\documentstyle[11pt]{article}

\begin{document}
\title{Steady and Oscillatory Side-Band Instabilities in Marangoni Convection
       with Deformable Interface}
\author{\dag A.A.Golovin, \ddag A.A.Nepomnyashchy,
          \dag L.M.Pismen, and $^{*}$H.Riecke \\
\\ \dag Department of Chemical Engineering, \\
   \ddag Department of Mathematics, \\
 Minerva Centre for Nonlinear Physics of Complex Systems, \\
 Technion, Haifa 32000, Israel; \\
  $^{*}$Department of Engineering Sciences and Applied Mathematics, \\
   Northwestern University, Evanston, USA.}
\maketitle
\begin{abstract}
The stability of Marangoni roll convection in a liquid-gas system with
deformable interface is studied in the case when there is a nonlinear
interaction between two modes of Marangoni instability: long-scale surface
deformations and short-scale convection. Within the framework of a model
derived in \cite{GNP}, it is shown that the nonlinear interaction between
the two modes substantially changes the width of the band of stable wave
numbers of the short-scale convection pattern as well as the type of the
instability limiting the band.  Depending on the parameters of the system,
the instability can be either long- or short-wave, either monotonic or
oscillatory. The stability boundaries strongly differ from the standard
ones and sometimes exclude the band center. The long-wave limit of the
side-band instability is studied in detail within the framework of the
phase approximation. It is shown that the monotonic instability is always
subcritical, while the long-wave oscillatory instability can be
supercritical, leading to the formation of either travelling or standing
waves modulating the pattern wavelength and the surface deformation. 
\end{abstract}
\newpage
\baselineskip=18pt
\section{Introduction.}

An important class of instabilities of one-dimensional patterns are
side-band instabilities. The most important one among them is the Eckhaus
instability \cite{Ech} which is a monotonic long-wave instability that
renders the pattern unstable with respect to infinitesimal compressions
and dilatations. Its onset can be characterized by the vanishing phase
diffusion coefficient \cite{CF1}. Near the threshold, where the pattern
can be described by a real (dissipative) Ginzburg--Landau (GL) equation,
the Eckhaus instability arises generically if the deviation $q$ of the
wave-number of the pattern from that corresponding to the bifurcation
point exceeds a certain value, i.e. if
\begin{equation} 
        |q|>\epsilon/\sqrt{3}, \label{ech} 
\end{equation} 
where $\epsilon^{2}$ is the deviation of the control parameter from the
threshold.  A similar instability of wave patterns (e.g. described  
by the complex GL equation) is usually referred to as the Benjamin-Feir
instability \cite{LK}. Typically, the Eckhaus instability restricts the
range of allowed wave numbers of the pattern, i.e. its nonlinear evolution
eventually brings the system back to the stable range of wave numbers. 

The Eckhaus instability of periodic patterns has been studied extensively,
both theoretically and experimentally, in various pattern-forming systems,
such as Rayleigh-B\'enard convection \cite{RB1,RB2}, convection in binary
mixtures \cite{BM}, Taylor vortex flow \cite{TW,RP}, electrohydrodynamic
convection in nematics \cite{NM}, directional solidification \cite{DS}.
Even quite close to the threshold, the boundaries of the Eckhaus
instability of stationary periodic patterns can differ from (\ref{ech})
due to various factors, such as the system confinement \cite{CF1,CF2},
subcritical character of the instability \cite{SB}, the shape of the
neutral stability curve \cite{NSC,RR}, the effect of nonlinearity
\cite{NL} and resonances \cite{RP}, the presence of random fluctuations
\cite{RF}, chiral symmetry breaking (e.g.due to rotation) \cite{RW}, etc. 

Although the Eckhaus instability is the most common one-dimensional
side-band instability, others are possible as well. In the present paper
we show that spatially periodic patterns arising due to Marangoni convection
in a liquid layer with a deformable free surface heated from
below\footnote{This type of convection can be also driven by
mass rather than heat transfer \cite{GNP}} can possess additional
side-band instabilities: a short-scale steady side-band instability and a
long-scale oscillatory side-band instability. An important feature of this
system is the presence of {\em two} primary modes of instability
(generating the Marangoni convection itself): a long-scale instability
(with zero wave-number) connected with the deformation of the free
surface, and short-scale convection (with a finite wave-number) governed
by surface tension gradients. A typical neutral stability curve for this
system has two minima, corresponding to {\em two} critical Marangoni
numbers \cite{Tksh}. The one minimum, Ma$_{l}$, is at a zero wave-number
and corresponds to a long-scale instability driven by deformation of the
free surface coupled with surface tension gradients. The other
minimum, Ma$_{s}$, is at a non-zero wave-number, and corresponds to a
short-scale instability caused by surface tension gradients only (surface
deformations are not important in this regime). 

When the two thresholds are close to each other, the nonlinear interaction
between the two modes of Marangoni instability with different scales can
lead to various secondary instabilities of the short-scale convection
generating various stationary and temporally modulated patterns
\cite{GNP,KSGP}. The nonlinear evolution and interaction of the two modes
is described by a system of coupled equations which can be written in the
following scaled form \cite{GNP}: 
\begin{eqnarray}
A_{t}&=& A+A_{xx}-|A|^{2}A+Ah,   \label{ah}  \\
h_{t}&=&-mh_{xx}-wh_{xxxx}+s|A|^{2}_{xx}. \nonumber
\end{eqnarray}
Here $A$ is the amplitude of the short-scale convection, $h$ is the
deformation of the free surface, $x$ and $t$ are long-scale space
coordinate and slow time. The form of the equation for $h$ is determined
by the fact that $h$ is conserved due to mass conservation.  In the
non-conserved case certain aspects of the interaction between short- and
long-scale structures has been discussed in \cite{MP}. The parameter $m$
is proportional to the difference between the two thresholds,
$\mbox{Ma}_{s}-\mbox{Ma}_{l},\;\;w>0$ characterizes the effect of the
Laplace pressure, and $s>0$ is the computable coupling parameter which was
shown to be positive \cite{GNP}. Note that for the terms in the second
equation in (\ref{ah}) to be of the same order, the capillary number,
which parameter $w$ is proportional to, has to be large (which is indeed
the case in most of experimental conditions).  The coupling between $A$
and $h$ reflects the following mechanism of the modes interaction:  as
expressed by the coupling term in the first equation, the short-scale
convection is more intensive beneath surface elevations; since $s>0$, the
convection, in turn, suppresses the long-scale deformational mode by
changing the average surface temperature or concentration \cite{GNP}. This
stabilization effect of the short-scale convection mode allows the
difference between the critical Marangoni numbers Ma$_{l}$ and Ma$_{s}$ to
be ${\cal O}(1)$ and still the deformational instability to be restricted
to long scales and have small growth rate. However, this stabilization
fails if Ma$_{s}-$Ma$_{l}$ is too large (see \cite{KSGP}). 

In Ref.\cite{GNP}, the effect of the mode interaction on the stability
of the short-scale convection was studied only in the vicinity of the
band center ($A=1,\;h=0$). In the present paper we extend the analysis to 
the full wave-number band of the short-scale convection and show that the
interaction of the two modes strongly affects the width of the band and can 
even lead to a change of the type of the instability.

\section{Linear Stability Analysis of Short-Scale Convection Rolls}

We consider the effect of surface deformations on the stability of
short-scale convection rolls with wave-numbers lying within an ${\cal
O}(\epsilon)$ range at small supercriticality
(Ma$-$Ma$_{s}$)/Ma$_{s}\equiv \epsilon^{2}$.  This convection pattern is
described by the following stationary solution of eq.(\ref{ah}): 
\begin{equation}
   A_{0}=\sqrt{1-q^{2}}\,e^{iqx},\;h=0,       \label{Aq}
\end{equation}
where $q<1$ is the dimensionless (in units of the characteristic width
$\epsilon$ of the band of excited wave-numbers) deviation from the
critical wave-number corresponding to the threshold Ma=Ma$_{s}$ of the
short-scale instability. 

Consider a perturbed solution in the form
\begin{eqnarray}
A&=&A_{0}(1+ae^{\sigma t+i\kappa x}+
                  be^{\sigma^{*} t-i\kappa x}), \label{ptrb} \\
h&=&ce^{\sigma t+i\kappa x}+c^{*}e^{\sigma^{*} t-i\kappa x}, \nonumber
\end{eqnarray}
where asterisks denote complex conjugates.
Substituting (\ref{ptrb}) in system (\ref{ah}) yields the following
implicit dispersion relation for the side-band instability:
\begin{equation}
\sigma^{3}+\alpha\sigma^{2}+\beta\sigma+\gamma=0, \label{disp}
\end{equation}
where
\begin{eqnarray*}
\alpha&=&-m\kappa^{2}+w\kappa^{4}+2(1-q^{2}+\kappa^{2}),\\
\beta&=&2(-m\kappa^{2}+w\kappa^{4})(1-q^{2}+\kappa^{2})+
        \kappa^{4}+2\kappa^{2}(1-3q^{2})+2s\kappa^{2}(1-q^{2}),\\
\gamma&=&(-m\kappa^{2}+w\kappa^{4})(\kappa^{4}+2\kappa^{2}(1-3q^{2}))+
                                           2s\kappa^{4}(1-q^{2}).
\end{eqnarray*}
We see that the interaction of the two modes leads to a {\em cubic} equation
for the eigenvalues defining the limits of the side-band instability
(similar, say, to the case of the Eckhaus instability of a hexagonal
pattern when there is a coupling between three waves \cite{Alik,Tsmr}). In
our case, we have the coupling between two side-band modes, which
correspond to longitudinal compression-dilatation waves propagating in
opposite directions if $\sigma$ is complex, and the surface deformation. 

According to the Routh-Hurwitz criterion, it follows from (\ref{disp})
that the stationary solution (\ref{Aq}) is stable if
\begin{equation}
          \alpha>0,\; \gamma>0,\; \alpha\beta-\gamma>0. \label{cond}
\end{equation}

The analysis of conditions (\ref{cond}) shows that, depending on the
values of the parameters $m,\;w\;\mbox{and}\;s$, the side-band instability
in this system can be either long-wave or short-wave, monotonic or
oscillatory, and the range of $q$ where the solution (\ref{Aq}) is stable
can be either wider or narrower than the standard Eckhaus interval defined
by (\ref{ech}). 

A detailed analysis of various possibilities is very complicated since
the stability of the system is governed by four parameters,
$m,\;w,\;s,\;\mbox{and}\;q$. We shall dwell therefore mostly on the 
case when the solution (\ref{Aq}) becomes unstable with respect to
perturbations with infinitesimal wave-number. The corresponding stability
limits can be determined from (\ref{cond}) in the limit $\kappa
\rightarrow 0$. It is, however, more instructive to obtain this result
using the so-called {\em phase approximation} which is based on the
observation that in the long-wave limit the evolution of the amplitude of
the convection pattern is slaved to the evolution of its phase
\cite{CN,CH}. Thus, we set $A=(\rho_{0}+\epsilon\rho)\exp{(iqx+\phi)},\;
h={\cal O}(\epsilon),\;\partial_{x}={\cal
O}(\epsilon),\;\partial_{t}={\cal O}(\epsilon^{2})$, and obtain from
eq.(\ref{ah}) for the amplitude correction $\rho$
\begin{equation}
    \rho=-\frac{q}{\rho_{0}}\phi_{x}+\frac{h}{2\rho_{0}},\;\; 
    \rho_{0}=\sqrt{1-q^{2}}.       \label{rofi}
\end{equation}
In our case, the amplitude near the instability threshold
is slaved to both the phase $\phi$ and the surface deformation $h$.
The evolution of these two variables is described by the following
system of coupled equations:
\begin{eqnarray}
   \phi_{t}&=&\frac{1-3q^2}{1-q^2}\phi_{xx}+
                          \frac{q}{1-q^2}h_{x}, \label{fih} \\
   h_{t}&=&(s-m)h_{xx}-2qs\phi_{xxx}. \nonumber
\end{eqnarray}
Introducing the wave-number deviation, $k \equiv \phi_{x}$, one can rewrite system
(\ref{fih}) in the form of the coupled diffusion equations:
\begin{equation}
    {\bf a}_{t}=-S{\bf a}_{xx},         \label{aS}
\end{equation}
where
\[ {\bf a}=\left( \begin{array}{c} k \\ h \end{array} \right), \;
   S=\left(\begin{array}{cc}
                \frac{3q^{2}-1}{1-q^2} & -\frac{q}{1-q^{2}} \\
                2qs  &   m-s \end{array} \right). \\
\]
The system is stable if all eigenvalues of matrix $S$ have negative
real parts. The characteristic equation leads to the following stability
conditions:
\begin{eqnarray}
    3q^{2}-1+(m-s)(1-q^2)<0, \label{stab1} \\
    (3q^{2}-1)(m-s)+2q^{2}s>0. \label{stab2}
\end{eqnarray}
 
It should be noted that this system obtained in the long-wave limit does not 
give the complete stability information. It can be shown that there always
exists an interval of $m$ where the relevant instability is a {\em
short-wave} instability that can be either monotonic or oscillatory. 

We shall dwell on the case when the instability is long-wave in the band
center, and no short-wave {\em oscillatory} instability arises.  A typical
stability diagram is shown in Fig.1 for $s=3,\;w=15$. Fig.1 is true
provided $\{s<2w,\;s<1+\sqrt{8w}\}$ (see Ref.\cite{GNP}) and $w$ is larger
than a certain value $w^{*}(s)$, respectively (for $s=3,\;w>w^{*}\approx
8)$. Fig.1 presents the region in the plane $(m,q^{2})$ in which the
solution (\ref{Aq}) is stable. This region is bounded by the lines AB, BC,
CD and DE. The lines AB and DE are the loci of $q^2=(m-s)/(3m-s)$ and
correspond to a {\em monotonic} long-wave instability (cf. (\ref{stab2})),
while the line CD is the locus of $q^2=(1+s-m)/(3+s-m)$ and corresponds to
an {\em oscillatory} long-wave instability (cf. (\ref{stab1})). The line
BC corresponds to a monotonic {\em short-wave} instability and can be
found analytically from the condition $\gamma=0$ (see Eq.(\ref{cond})) (it
is, however, too cumbersome to be reproduced here). The horizontal dashed
line corresponds to the threshold of the standard Eckhaus instability,
$q^{2}=1/3$. 

The coordinates of the points A, B, C, D, and E can be found analytically:
A$\rightarrow \{-\infty,1/3\}$,
B$=\{(s-s_{w})/6,\;(5s+s_{w})/(3(s+s_{w}))\},\;s_{w}=\sqrt{s^2+48sw}$,
D$=\{s(1+\sqrt{1+4/s})/2,\; (\sqrt{1+4/s}-1)/(3\sqrt{1+4/s}+1)\}$,
E$=\{s,0\}$; the expressions for the coordinates of the point C are very
lengthy (for $s=3,\;w=15$ in Fig.1, C$=\{1.97,\;0.5\}$). 

Thus, due to the coupling between the phase evolution and the surface
deformation, both the interval of wave-numbers corresponding to stable
stationary convection described by eq.(\ref{Aq}) and the type of its
instability depend on the parameter $m$, i.e. on the difference between
the two thresholds of the Marangoni instability, Ma$_{s}-$Ma$_{l}$. There
are four particular values of
$m\;(m_{1}(s,w),\;m_{2}(s,w),\;m_{3}(s),\;m_{4}(s))$, corresponding to the
points B, C, E, D in Fig.1, respectively, which define four regions where
qualitatively different instabilities are observed. These regions in the
$(s,m)$-parameter plane are shown in Fig.2 for $w=15$. 

If $m<m_{1}(s,w)$, solution (\ref{Aq}) is stable when $q^{2}<(m-s)/(3m-s)$.
Since $s>0$ in our case, the stability interval for $q$ is {\em larger}
than in the case of the ordinary Eckhaus instability, $q^{2}<1/3$; part of
the energy of phase perturbations is transferred to the damped long-scale
deformational mode. This suppresses the Eckhaus instability and makes the
stability interval for $q$ larger. Outside this interval of $q$, solution
(\ref{Aq}) is unstable and the instability is {\em monotonic} and {\em
long-wave}, like in the standard case with no surface deformation. In the
limit $m\rightarrow -\infty$, surface deformations are damped very
strongly, and the Eckhaus interval tends to its ordinary value. 

When $m_{1}(s,w)<m<m_{2}(s,w)$, the stability threshold for $q$ is
determined by a cumbersome function of $w$ and $s$ which we do not
represent here. This threshold is still larger than $1/3$. The instability
in this case is also monotonic, but {\em short-wave}. At the point B the
modulation wave number $\kappa$ goes to zero and the short-wave
instability continuously changes over to a long-wave instability. This
shows that the short-wave instability is not due to a subharmonic
resonance as it has been found, e.g., in Taylor vortex flow \cite{PR}. 

When $m_{2}(s,w)<m<m_{3}(s)=s$, the stability interval for $q$ is given by
$q^{2}<(1+s-m)/(3+s-m)$. The instability in this case is {\em oscillatory}
and {\em long-wave} and can lead to slow oscillations of the wave-number
of the short-scale roll pattern coupled with oscillations of the surface
deformation (see sec.3.2 below). 

At $m=m_{3}(s)=s$, the stability region coincides with the ordinary
Eckhaus interval, $q^{2}<1/3$.

If $m>m_{3}(s)=s$, the roll pattern corresponding to the band center, $q=0$,
becomes unstable. This coincides with the results in \cite{GNP}.
Nevertheless, as long as $m_{3}(s)<m<m_{4}(s)$, a {\em stable} pattern
{\em still exists} under these conditions, but with a {\em modified}
wave-number lying in the interval $(m-s)/(3m-s)<q^{2}<(1+s-m)/(3+s-m)$.
The lower boundary of this interval corresponds to the long-wave {\em
monotonic} instability (i.e. the usual Eckhaus instability) while the
upper one to the long-wave {\em oscillatory} instability. This splitting
of the stable band into two sub-bands is reminiscent of a situation
encountered in parametrically excited standing waves: under certain
conditions the wave numbers in the center of the band become
Eckhaus-unstable while off-center wave numbers remain stable.  Such a
splitting was found to lead to the formation of stable kinks in the wave
number separating domains with different wave numbers \cite{RK1}. The
nonlinear evolution of the present system in this regime will be discussed
briefly below. 

For $m>m_{4}(s)$, solution (\ref{Aq}) is unstable at any wave number due to
the deformational instability of the surface. 

Thus, depending on $m$, i.e. on the difference between the threshold
Ma$_{s}$ of the short-scale convection and that of the long-scale
deformational instability (Ma$_{l}$), the band of stable wave numbers can
take on two qualitatively different forms. For $m<s$ the band is
contiguous (hashed area in Fig.3a) while for $s<m<m_{4}(s)$ convection is
stable within two disconnected wave-number bands (hashed area in Fig.3b). 

In conclusion of this section let us note that if the parameter $w$ is
{\em not} large enough, the {\em oscillatory short-wave} side-band
instability comes into play. In this case, the stability region shown in
Fig.1 slightly changes: there appears a new boundary corresponding to this
short-wave oscillatory side-band instability which cuts off the corner at
the point C and connects two boundaries BC and CD corresponding to the
short-wave monotonic and to the long-scale oscillatory side-band
instabilities, respectively. 

\section{Nonlinear evolution of the long-wave instabilities}

In this section we shall study the nonlinear evolution of the long-wave
instability near the thresholds given by the curves AB, CD and DE in
Fig.1. 

\subsection{Eckhaus instability}

The monotonic long-wave instability threshold corresponds to the
curves AB and DE in Fig.1 and is given by
\begin{equation}
        q_{m}^{2}=\frac{m-s}{3m-s}. \label{em}
\end{equation}
We use the following scaling: $X=\epsilon x,\;T=\epsilon^{4}t$, and apply
the following ansatz:
\begin{eqnarray}
A=\rho\exp{i\phi}, \label{Arofi} \\
\rho=\rho_{0}+\epsilon^{2}\rho_{1}(X,T)+
           \epsilon^{3}(X,T)\rho_{2}+\ldots, \nonumber \\
\phi=qx+\epsilon\phi_{1}(X,T)+\epsilon^{2}\phi_{2}(X,T)+\ldots, \nonumber \\
h=\epsilon^{2}h_{1}(X,T)+\epsilon^{3}h_{2}(X,T)+\ldots, \nonumber \\
q=q_{m}+\epsilon^{2}q_{2}+\ldots . \nonumber
\end{eqnarray}

Substituting (\ref{Arofi}) in system (\ref{ah}), one obtains in the first
order that the change in the local wave-number, $k_{1}\equiv
\partial_{X}\phi_{1}$, is proportional to the surface deformation $h_{1}$,
\begin{equation}
      k_{1}=\frac{s-m}{2sq_{m}}h_{1}. \label{k1h1m}
\end{equation}
Proceeding to the sixth order, we  obtain the following
nonlinear evolution equation for the surface deformation $h_{1}$:
\begin{equation}
 \partial_{T}h_{1}+\sigma_{1}\partial_{XX}h_{1}+
    \sigma_{2}\partial_{XXXX}h_{1}+
            \sigma_{3}\partial_{XX}(h_{1}^{2})=0, \label{evh} 
\end{equation}
where
\begin{eqnarray*}
\sigma_{1}=q_{2}q_{m}\frac{(4m-s)(s-3m)}{2(-m^2+ms+s)},\;
\sigma_{2}=\frac{-3m^2+ms+4sw}{4(-m^2+ms+s)},\\
\sigma_{3}=\frac{m(s-3m)(5s-6m)}{8s(-m^2+ms+s)}. 
\end{eqnarray*}

The evolution equation for the long-wave modulation of the wave-number
$k_{1}$ can be obtained from eq.(\ref{evh}) by a simple renormalization of
the coefficient $\sigma_{3}\rightarrow\sigma_{3}2sq_{m}/(s-m)$ (cf.
(\ref{k1h1m})).  When $m\rightarrow -\infty$, the surface is undeformable,
and the coefficients in the equation for $k_{1}$ tend to those of the
nonlinear evolution equation describing an ordinary Eckhaus instability in
the real GL equation. 

It can be shown that the coefficients $\sigma_{2}$ and $\sigma_{3}$ are
positive (in the parameter region where the instability is long-wave), and
the instability is subcritical. Therefore, all periodic solutions of
(\ref{evh}) blow up beyond the stability limit $\sigma_1=0$ indicating a
break-down of (\ref{evh}). The same happens in the case of the {\em
monotonic short-scale} instability, corresponding to the curve BC in
Fig.1; our computations show that it is also subcritical.  However,
numerical simulations of the full amplitude equations (\ref{ah}) show that
different physical processes occur when crossing the lines AC and DE in
Fig.1, respectively. This is related to the fact that along AC the
deformational mode is linearly damped ($m<s$), whereas along DE it is
excited ($m>s$).  Figs.4,5 show typical results of numerical simulations
of (\ref{ah}). 

Fig.4 shows the evolution of the pattern occurring when the line AC is
crossed. While the surface deformation and the amplitude of the
short-scale convection remain regular for all times, the wave-number goes
through a true singularity. It signifies a phase slip, which reduces the
total phase in the system by $2\pi$. Following this phase slip, the
pattern relaxes to a periodic pattern with a wave-number in the stable
band. 

The evolution below the line DE is shown in Fig.5. Here the deformational
mode is excited; it has been shown previously that the instability is
subcritical at $q=0$ \cite{KSGP}. The simulation of the coupled amplitude
equations (\ref{ah}) indicates that the instability does indeed not
saturate which results in a singularity in the surface deformation and in
the amplitude of the short-scale convection.  Thus, even the amplitude
equations (\ref{ah}) become invalid. This result suggests a dry-out of the
fluid film as has been observed experimentally \cite{Sw}. 

Despite the splitting of the wave-number band, no stable wave-number kinks
could be found. This indicates that the higher-order contributions to the
phase equation (\ref{evh}), which were not calculated, differ from those
arising solely from a non-monotonicity of the phase diffusion coefficient
\cite{RK1}. 

It should be noted that at the point D the coefficients
$\sigma_{1},\;\sigma_{2}$ and $\sigma_{3}$ tend to infinity, while at
the point E the coefficient $\sigma_{1}$ vanishes regardless of the
value of $q_{2}$; the system dynamics in the vicinity of these points is
more complex; it cannot be described by eq.(\ref{evh}) and requires
separate analysis.

\subsection{Oscillatory long-wave instability}

Along line CD in Fig.1 an oscillatory long-wave instability arises.
Its threshold is given by
\begin{equation}
   q_{o}^{2}=\frac{1+s-m}{3+s-m}. \label{eo}
\end{equation}
To study its nonlinear behavior we set $A=\rho\exp{i\phi}$, introduce
 the slow coordinates
$X=\epsilon x$, $T_{0}=\epsilon^{2}t,\;
T=\epsilon^{4}t$, and expand
\begin{eqnarray}
\rho=\rho_{0}+\epsilon\rho_{1}(X,T_{0},T)+
           \epsilon^{2}(X,T_{0},T)\rho_{2}+
             \epsilon^{3}(X,T_{0},T)\rho_{3}+\ldots, \label{expo} \\
\phi=qx+\phi_{0}(X,T_{0},T)+\epsilon\phi_{1}(X,T_{0},T)+
             \epsilon^{2}\phi_{2}(X,T_{0},T)+
                \epsilon^{3}\phi_{3}(X,T_{0},T)+\ldots, \nonumber \\
h=\epsilon h_{1}(X,T_{0},T)+\epsilon^{2}h_{2}(X,T_{0},T)+
                   \epsilon^{3}h_{3}(X,T_{0},T)+\ldots, \nonumber \\
q=q_{o}+\epsilon^{2}q_{2}+\ldots,\; \nonumber
\end{eqnarray}
with
 \[
                         \rho_{0}=\sqrt{1-q_{o}^2}. 
\]
Substituting (\ref{expo}) in system (\ref{ah}), we obtain in the first order
a linear system for $h_{1}$ and $k_{0}=\partial_{X}\phi_{0}$ which, after
appropriate diagonalization, can be reduced to a linear free Schr\"{o}dinger
equation
\begin{equation}
  i\partial_{T_{0}}\psi=-\omega_{0}\partial_{XX}\psi, \label{schr}
\end{equation}
where
\[
  \psi=((m-s+i\omega_{0})h_{1}-k_{0})/(2i\omega_{0}),\;
       \omega_{0}=\sqrt{-m^2+s(m+1)}.
\]
Eq.(\ref{schr}) describes waves propagating along the convection pattern
and modulating its wave-number and the surface deformation. The phase
velocity of the waves $v_{p}$ depends on the wave number
$\kappa$ of the modulation, as $v_{p}=\omega_{0}\kappa$. The wave-number 
$k_{0}$ and the surface deformation $h_{1}$ are proportional to each other
\begin{equation}
  k_{0}=\delta h_{1},\; \delta=\frac{s-m+i\omega_{0}}{2sq_{0}}, \label{k0h1}
\end{equation}
and there is a phase shift between oscillations of the wave-number and
the surface deformation equal to arg$(\delta)$.

At the onset of the oscillatory long-wave instability, a whole
band of wave-numbers $\sim O(\epsilon)$ is excited. However, a strong 
(quadratic) dispersion of the waves described by (\ref{schr}) 
allows only quartic interactions between waves with different
wave-numbers within the excited interval. The evolution of the wave
amplitudes can thus be described in the third order of the perturbation
theory by a system of Landau equations. A similar situation occured in a
study of long-scale oscillatory double-diffusive convection \cite{Psmn}.

In order to study the wave interaction, we consider two pairs of
oppositely propagating waves having wave-numbers $\kappa_{1}$ and
$\kappa_{2}$, and present $k_{0}$ and $h_{1}$ in the following
form:
\begin{eqnarray}
 \left(\begin{array}{c} k_{0}\\ h_{1} \end{array}\right)=
   \left(\begin{array}{c} \delta \\ 1 \end{array}\right)
  \left[A_{1}(T)e^{i\theta^{+}_{1}}+B_{1}(T)e^{i\theta^{-}_{1}}+
A_{2}(T)e^{i\theta^{+}_{2}}+B_{2}(T)e^{i\theta^{-}_{2}}\right]+c.c.,\label{AB}\\
 \theta^{\pm}_{1,2}=\omega_{0}\kappa_{1,2}^{2}T_{0}\pm\kappa_{1,2}X, \nonumber
 \end{eqnarray}
where the amplitudes $A_{1,2}$ and $B_{1,2}$ undergo slow evolution on the
time scale $T\equiv \epsilon^{4}t$. 

Using (\ref{AB}) and (\ref{expo}), we obtain from (\ref{ah}), as
solvability conditions in the fourth and in the fifth orders, the following
system of Landau equations for the amplitudes $A_{1,2}$ and $B_{1,2}$: 
\begin{eqnarray}
\dot{A_{1}}=\gamma_{1}A_{1}+
       \lambda^{a}_{11}|A_{1}|^{2}A_{1}+\lambda^{b}_{11}|B_{1}|^{2}A_{1}+
\lambda^{a}_{12}|A_{2}|^{2}A_{1}+
      \lambda^{b}_{12}|B_{2}|^{2}A_{1}, \label{land} \\
\dot{B_{1}}=\gamma_{1}B_{1}+
       \lambda^{a}_{11}|B_{1}|^{2}B_{1}+\lambda^{b}_{11}|A_{1}|^{2}B_{1}+
 \lambda^{a}_{12}|B_{2}|^{2}B_{1}+
    \lambda^{b}_{12}|A_{2}|^{2}B_{1}, \nonumber \\
\dot{A_{2}}=\gamma_{2}A_{2}+
       \lambda^{a}_{22}|A_{2}|^{2}A_{2}+\lambda^{b}_{22}|B_{2}|^{2}A_{2}+
 \lambda^{a}_{21}|A_{1}|^{2}A_{2}+
     \lambda^{b}_{21}|B_{1}|^{2}A_{2}, \nonumber \\
\dot{B_{2}}=\gamma_{2}B_{2}+
       \lambda^{a}_{22}|B_{2}|^{2}B_{2}+\lambda^{b}_{22}|A_{2}|^{2}B_{2}+
 \lambda^{a}_{21}|B_{1}|^{2}B_{2}+
      \lambda^{b}_{21}|A_{1}|^{2}B_{2}. \nonumber
\end{eqnarray}

The linear growth rate coefficients $\gamma_{n}$ and the Landau
coefficients $\lambda^{a,b}_{nn}\;(n=1,2)$ have the following form:
\begin{equation}
    \gamma_{n}=q_{2}\tilde{\alpha}(m,s)\kappa_{n}^{2}-
                   \tilde{\beta}(m,s,w)\kappa_{n}^{4},\;\;
    \lambda^{a,b}_{nn}=
         \tilde{\lambda}_{a,b}(m,s)\kappa_{n}^{2},\;\;n=1,2, \label{gamlam}
\end{equation}
where $\Re{(\tilde{\alpha}(m,s))}>0,\;\Re{(\tilde{\beta}(m,s))}>0$ if
$m_{2}<m<m_{4}$. The Landau coefficients $\lambda^{a,b}_{np}\;(n\neq p)$
are complicated functions of $m,\;s,\;\kappa_{1},\;\kappa_{2}$ satisfying
the relationships
\begin{eqnarray*}
 \lambda^{a,b}_{12}(\kappa_{1},\kappa_{2})=
 \lambda^{a,b}_{21}(\kappa_{2},\kappa_{1}),\;
 \lambda^{b}_{12}(\kappa_{1},\kappa_{1})=
 \lambda^{b}_{21}(\kappa_{1},\kappa_{1})=\lambda^{b}_{11},\\
 \lambda^{a}_{12}(\kappa_{1},\kappa_{1})=
                \lambda^{a}_{21}(\kappa_{1},\kappa_{1})
                       =f(m,s)\lambda^{a}_{11}.
\end{eqnarray*}

The real parts $\Re{(\tilde{\lambda}_{a,b}(m,s))}$ of the functions
$\tilde{\lambda}_{a,b}(m,s)$ in Eq.(\ref{gamlam}) are shown in Fig.6 as
functions of $m$. They become unbounded in the vicinity of the point D
($m=s(1+\sqrt{1+4/s})/2$) where system (\ref{land}) is not valid. 

System (\ref{land}) allows to study the competition between travelling and
standing waves. Let us consider two waves with a fixed wave-number propagating
in opposite directions. Setting $A_{2}\equiv B_{2}\equiv 0$ in (\ref{land}),
we study the stability of the two solutions of system (\ref{land})
\begin{eqnarray}
1)\;A_{1}^{2}&=&-\frac{\gamma_{1}}{\lambda^{a}_{11}},\;
           B_{1}=A_{2}=B_{2}=0\;\;\;\mbox{(travelling wave)}; \label{tw} \\
2)\;A_{1}^{2}&=&B_{1}^{2}=
    -\frac{\gamma_{1}}{\lambda^{a}_{11}+\lambda^{b}_{11}},\;
           A_{2}=B_{2}=0\;\;\;\mbox{(standing wave)}; \label{sw}
\end{eqnarray}
with respect to infinitesimal perturbations of the amplitudes $A_{1}$ and
$B_{1}$. The symbols of the Landau coefficients in (\ref{tw}), (\ref{sw})
and below denote now only the {\em real parts} of the respective
coefficients. 

The stability analysis shows that

a){\em travelling waves} are stable if
\begin{equation}
\lambda^{a}_{11}<0,\;
     \lambda^{b}_{11}<0,\;\lambda^{a}_{11}/\lambda^{b}_{11}<1; \label{t}
\end{equation}

b){\em standing waves} are stable if
\begin{equation}
\lambda^{a}_{11}<0,\;
                  |\lambda^{a}_{11}/\lambda^{b}_{11}|>1; \label{s}
\end{equation}
otherwise, the system undergoes a subcritical instability which may lead
to a blow-up and may return the system back to the stable region.  We have
performed some numerical simulations of (\ref{ah}) which support this
expectation. 

We have found (see Fig.6) that in our case ($w=15,\; s=3$), when
$2.135<m<3.361$ standing waves are stable, and when $3.361<m<3.745$
traveling waves are stable. In the rest intervals of $m$ within the
interval $m_{2}<m<m_{4}$, close to the points C and D, both types of waves
are subcritically unstable. 
  
Let us now study the stability of standing and travelling waves with
respect to perturbations with different wave-numbers, i.e. the stability
of solutions (\ref{tw}), (\ref{sw}) with respect to perturbations of the 
amplitudes $ A_{2}$ and $B_{2}$. The stability analysis shows that

a) {\em travelling waves} are stable if, besides (\ref{t}), the following
conditions are fulfilled:
\begin{equation}
\gamma_{2}-\gamma_{1}\frac{\lambda^{a}_{21}}{\lambda^{a}_{11}}<0,\;
 \gamma_{2}-\gamma_{1}\frac{\lambda^{b}_{21}}{\lambda^{a}_{11}}<0; \label{t1}
\end{equation}

b) {\em standing waves} are stable if, besides (\ref{s}), the following
condition holds:
\begin{equation}
 \gamma_{2}-\gamma_{1}\frac{\lambda^{a}_{21}+\lambda^{b}_{21}}
                           {\lambda^{a}_{11}+\lambda^{b}_{11}}<0; \label{s1}
\end{equation}

The analysis of conditions (\ref{t1}) and (\ref{s1}) shows that within the
intervals defined by (\ref{t}) and (\ref{s}) there exist intervals of
wave-numbers where the waves are also stable to perturbations with an
arbitrary wave-number. Fig.7 shows how these intervals change with the
variation of the parameter $m$ along curve CD in Fig.1. It can be seen that
standing waves are stable at intermediate wave-numbers, while travelling
waves can be stable at infinitesimal wave-number as well. 

Thus, the oscillatory long-wave Eckhaus instability in Marangoni
convection with deformable interface can generate either standing or
travelling compression-dilatation waves propagating along the periodic
pattern and modulating its wavelength together with the deformation of
the free surface. Typical results of numerical simulations of (\ref{ah})
are given in Fig.8a,b. It shows space-time diagrams 
for typical standing waves and traveling waves, respectively.

\section*{Acknowledgement.}
This research was supported by the Israel Science Foundation.
A.A.G. acknowledges the support of the Ministry for Immigrant
Absorption. H.R. gratefully acknowledges a travel grant by the Minerva Center and support
by DOE through DE-FG02-92ER14303.

\newpage

{\bf Figure 1.} The dependence of the wavenumber $q$ corresponding to the
onset of the various side-band instabilities on the parameter $m$
($w=15,\;s=3$). Different sections of the solid line correspond to
different types of the instability: AB -- long-wave monotonic (Eckhaus);
BC -- short-wave monotonic; CD -- long-wave oscillatory; DE -- long-wave
monotonic (Eckhaus). The dashed line presents the standard Eckhaus
instability threshold $q^{2}=1/3$. \\ \\
{\bf Figure 2.} Regions in the $(m,s)$-parameter plane corresponding to
different types of the side-band instability (at $w=15$): ML -- monotonic
long-wave (Eckhaus), MS -- monotonic short-wave, OL -- oscillatory
long-wave.\\ \\ 
{\bf Figure 3.} Two qualitatively different types of stable
wave-number bands for the short-scale Marangoni convection near the instability
threshold: \\
a) $m<m_{3}$: the stable region (hashed) is
larger than in the standard case (dashed line), and its boundary (solid
line) can correspond to either a monotonic, long- or short-wave
instability, or to an oscillatory long-wave instability; \\
b) $m_{3}<m<m_{4}$: the band center is unstable, and the stable
region (hashed) is split up into two parts; its inner boundary
corresponds to a monotonic long-wave instability and the outer one
to a long-wave oscillatory instability. The dotted line is the
neutral stability curve. \\ \\
{\bf Figure 4.} Typical space-time diagrams for the evolution of the
long-wave monotonic instability along line AC (see Fig.1) as obtained by
numerical simulation of (\ref{ah}): the surface deformation a) and the 
amplitude of the short-scale convection b) remain regular while the 
wavenumber passes through a singularity indicating the phase slip.\\ \\ 
{\bf Figure 5.} Typical space-time diagrams for the evolution of the
long-wave monotonic instability along line DE (see Fig.1) as obtained by
numerical simulation of (\ref{ah}). Here the surface deformation a) 
and the amplitude of the short-scale convection b) become
singular indicating dry-out of the fluid film.\\ \\
{\bf Figure 6.} The real parts of the interaction coefficients
$\Re{(\tilde{\lambda}_{a}(m))}$ (curve a) and
$\Re{(\tilde{\lambda}_{b}(m))}$ (curve b) from Eq.(\ref{gamlam}) at
$s=3,\;w=15$, as functions of $m$. The vertical asymptote corresponds to
$m=m_{4}(s)$ (point D in Fig.1). \\ \\
{\bf Figure 7.} Regions of stability of standing and travelling waves
with a wavenumber $\kappa_{1}$ (lying within the range of
linearly excited modes) with respect to perturbations  with a wavenumber
$kappa_{2}$ (lying within the same region), at different values of the
parameter $m$: s -- stable, u -- unstable.\\ \\
{\bf Figure 8.} a) Typical space-time diagram of a standing wave arising from
the long-wave oscillatory side-band instability of the periodic 
short-scale convection.\\
b) Typical space-time diagram of a traveling wave arising from
the long-wave oscillatory side-band instability of the periodic short-scale 
convection.\\
Both diagrams were obtained by numerical simulation of (\ref{ah}).

\newpage
\begin{figure}[htb]
\begin{picture}(420,240)(0,0)
\put(-100,-500) {\includegraphics{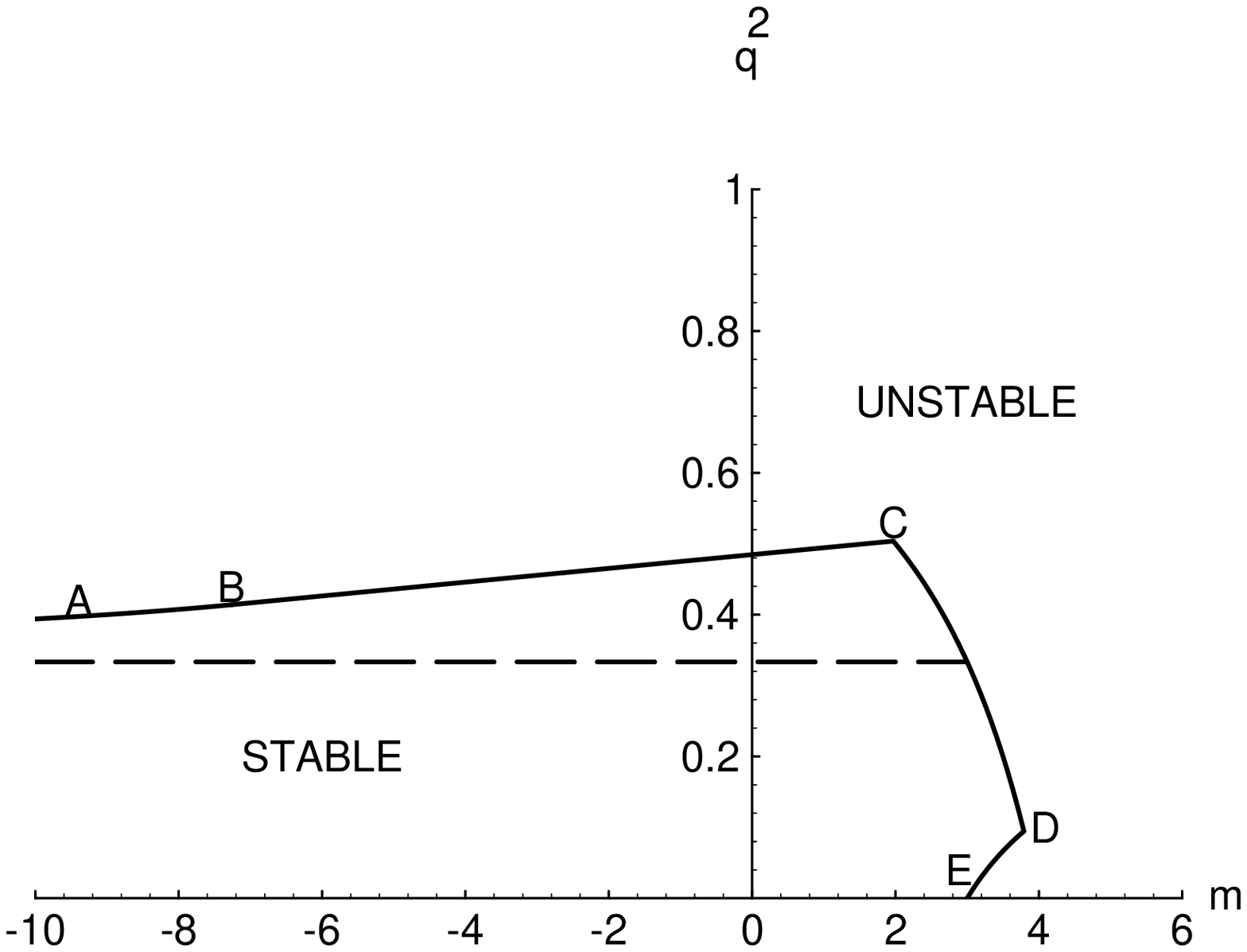}}
\end{picture}
\end{figure}

\newpage
\begin{figure}[htb]
\begin{picture}(420,240)(0,0)
\put(-100,-500) {\includegraphics{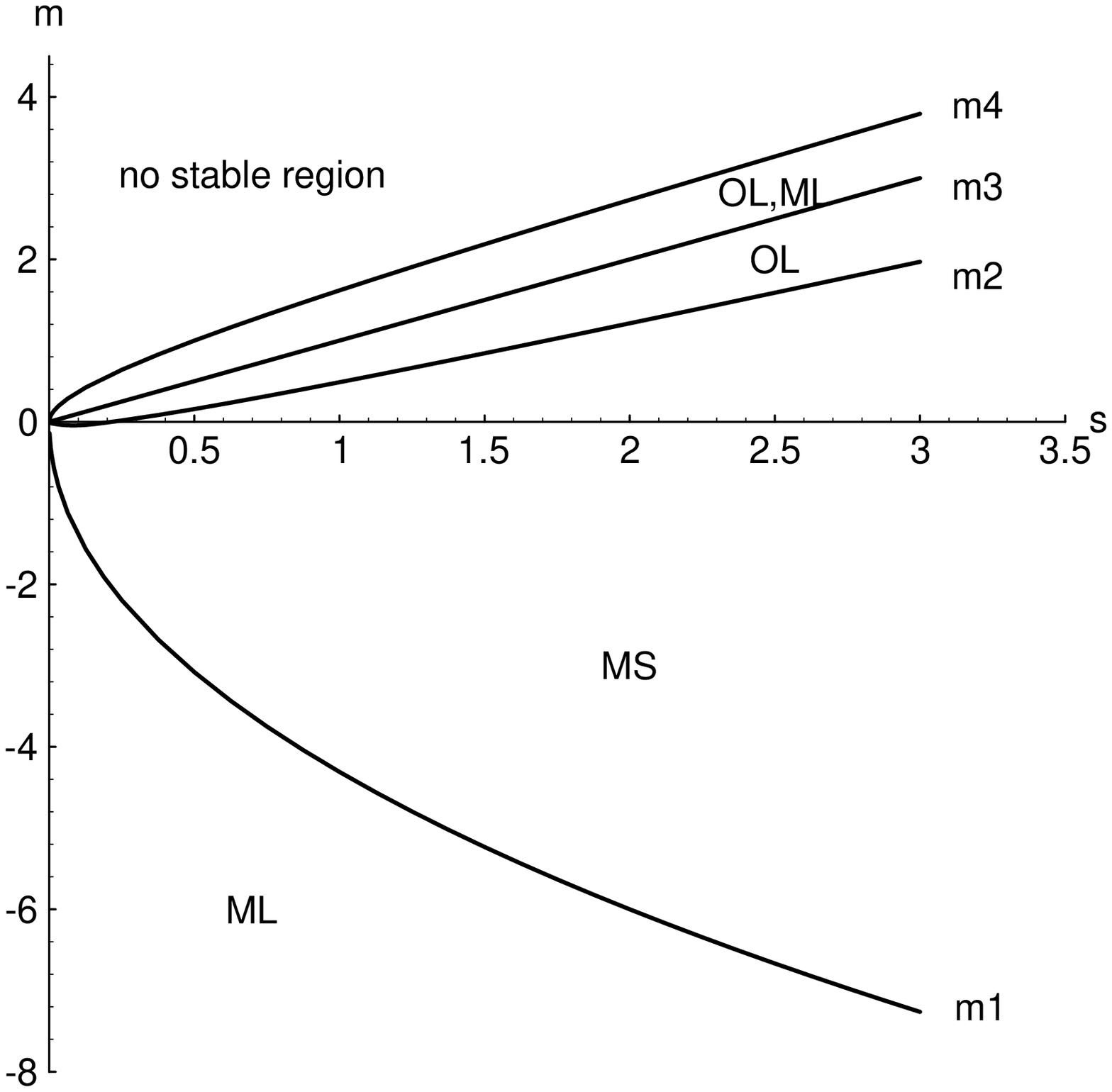}}
\end{picture}
\end{figure}

\newpage
\begin{figure}[htb]
\begin{picture}(420,240)(0,0)
\put(-100,-500) {\includegraphics{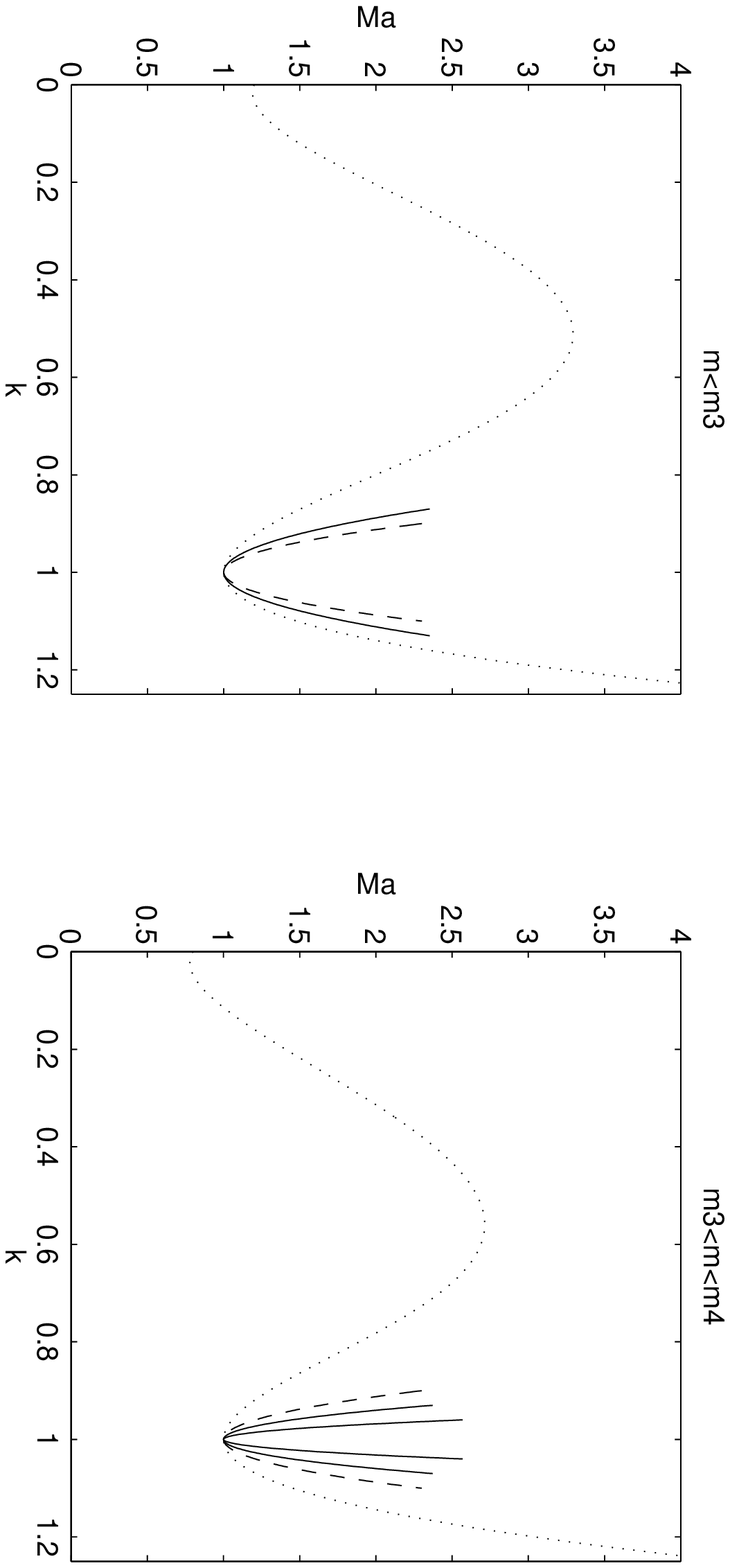}}
\end{picture}
\end{figure}

\newpage
\begin{figure}[htb]
\begin{picture}(420,240)(0,0)
\put(-100,-500) {\includegraphics{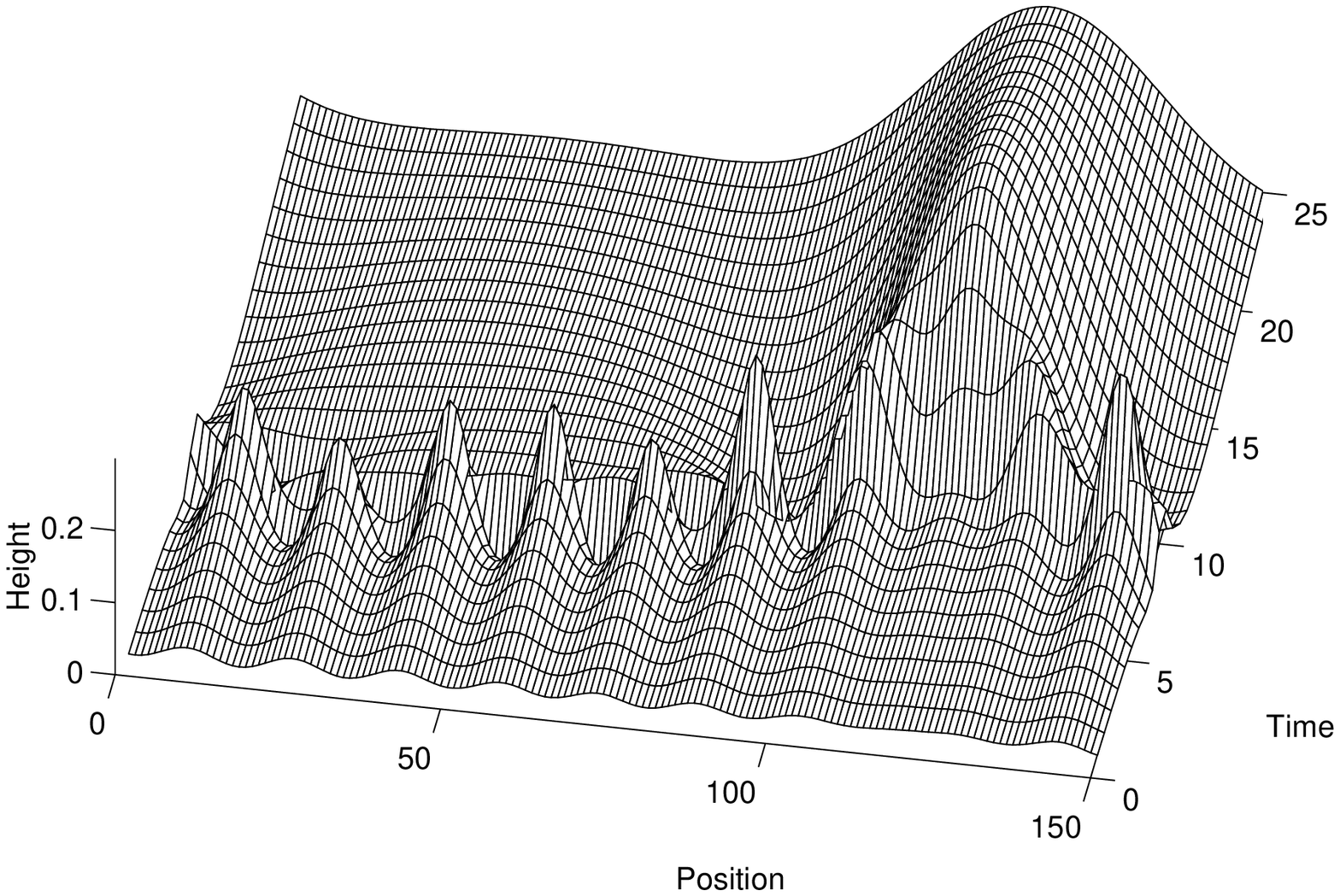}}
\end{picture}
\end{figure}

\newpage
\begin{figure}[htb]
\begin{picture}(420,240)(0,0)
\put(-100,-500) {\includegraphics{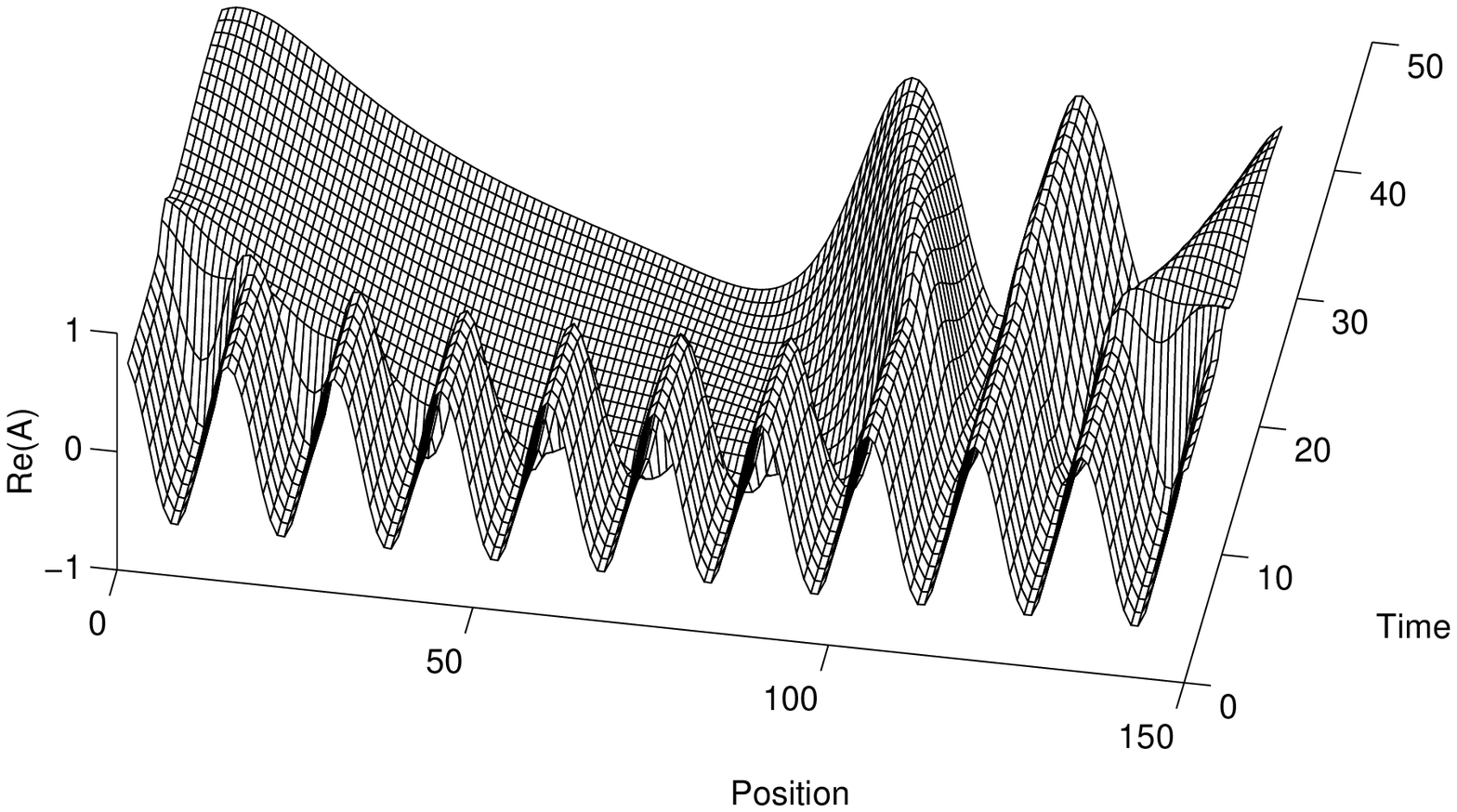}}
\end{picture}
\end{figure}

\newpage
\begin{figure}[htb]
\begin{picture}(420,240)(0,0)
\put(-100,-500) {\includegraphics{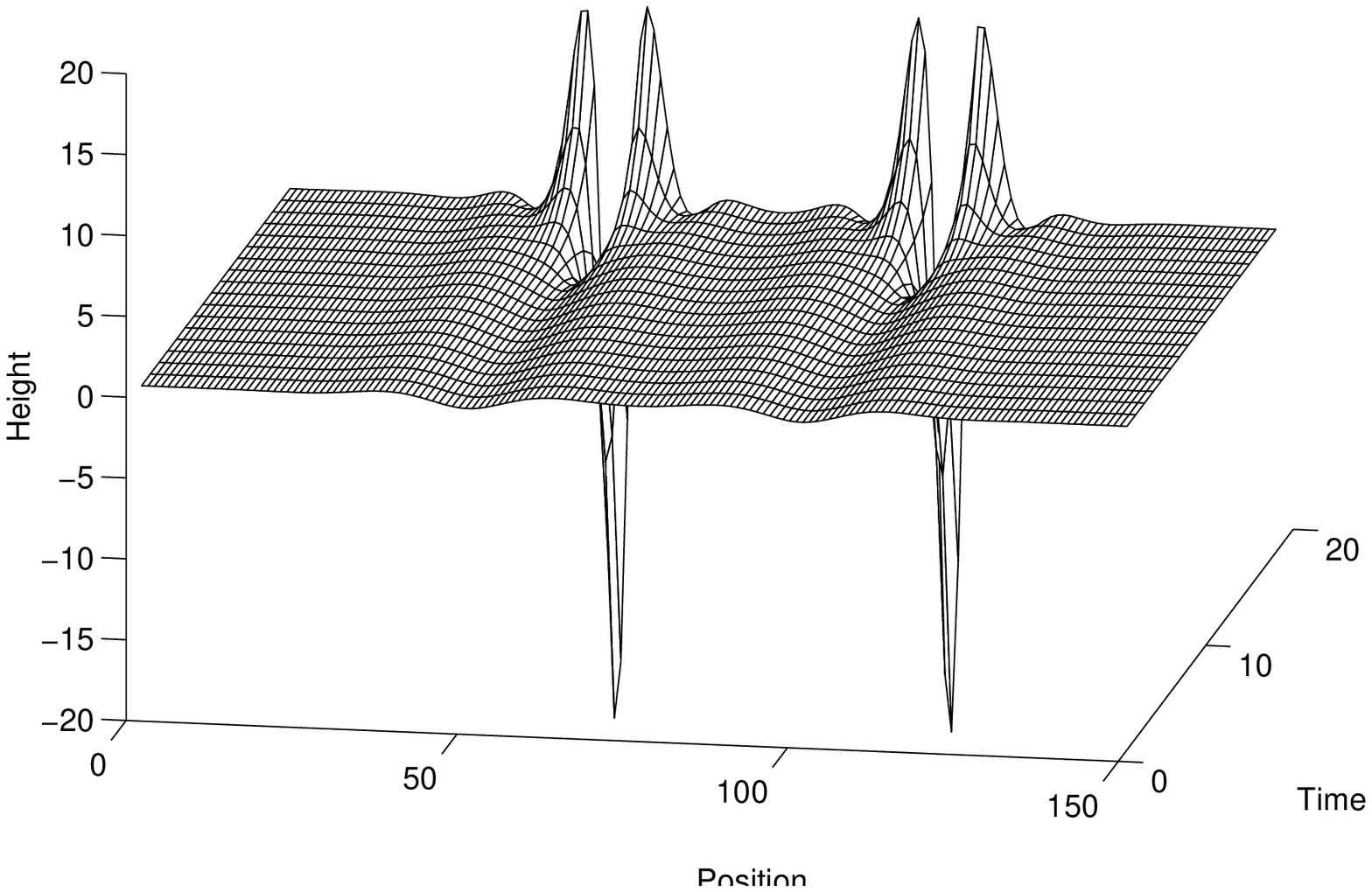}}
\end{picture}
\end{figure}

\newpage
\begin{figure}[htb]
\begin{picture}(420,240)(0,0)
\put(-100,-500) {\includegraphics{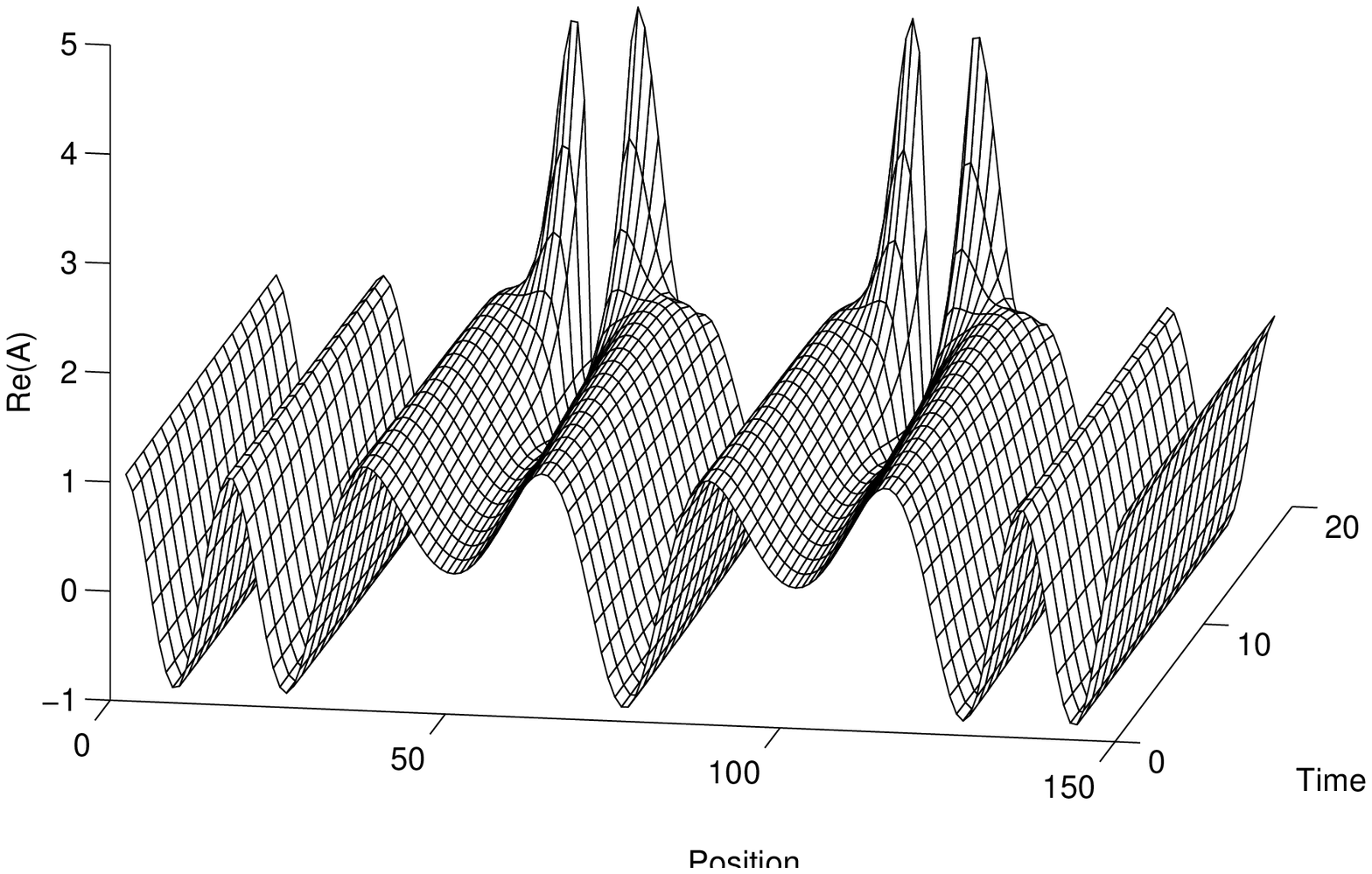}}
\end{picture}
\end{figure}

\newpage
\begin{figure}[htb]
\begin{picture}(420,240)(0,0)
\put(-100,-500) {\includegraphics{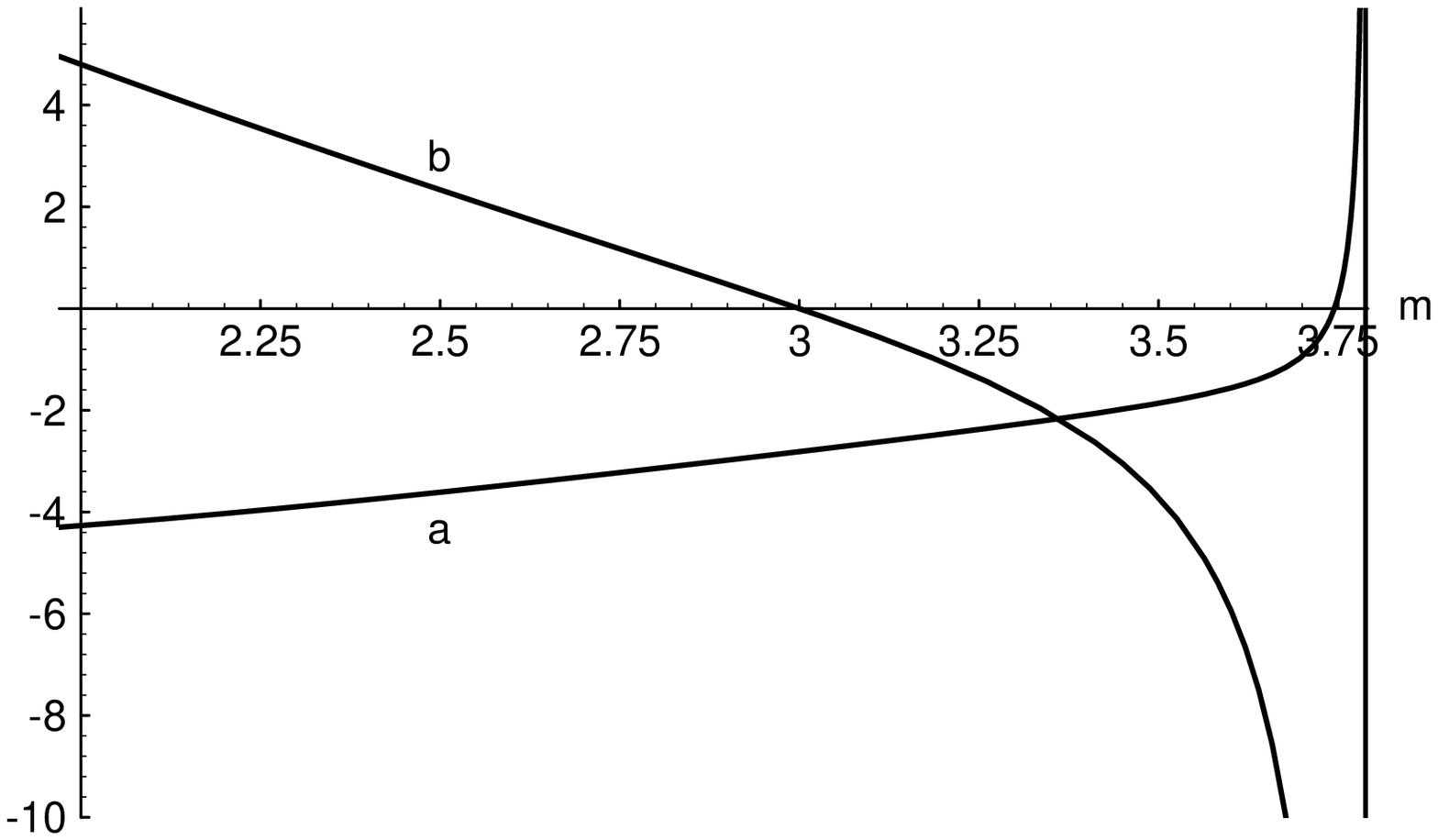}}
\end{picture}
\end{figure}

\newpage
\begin{figure}[htb]
\begin{picture}(420,240)(0,0)
\put(-100,-500) {\includegraphics{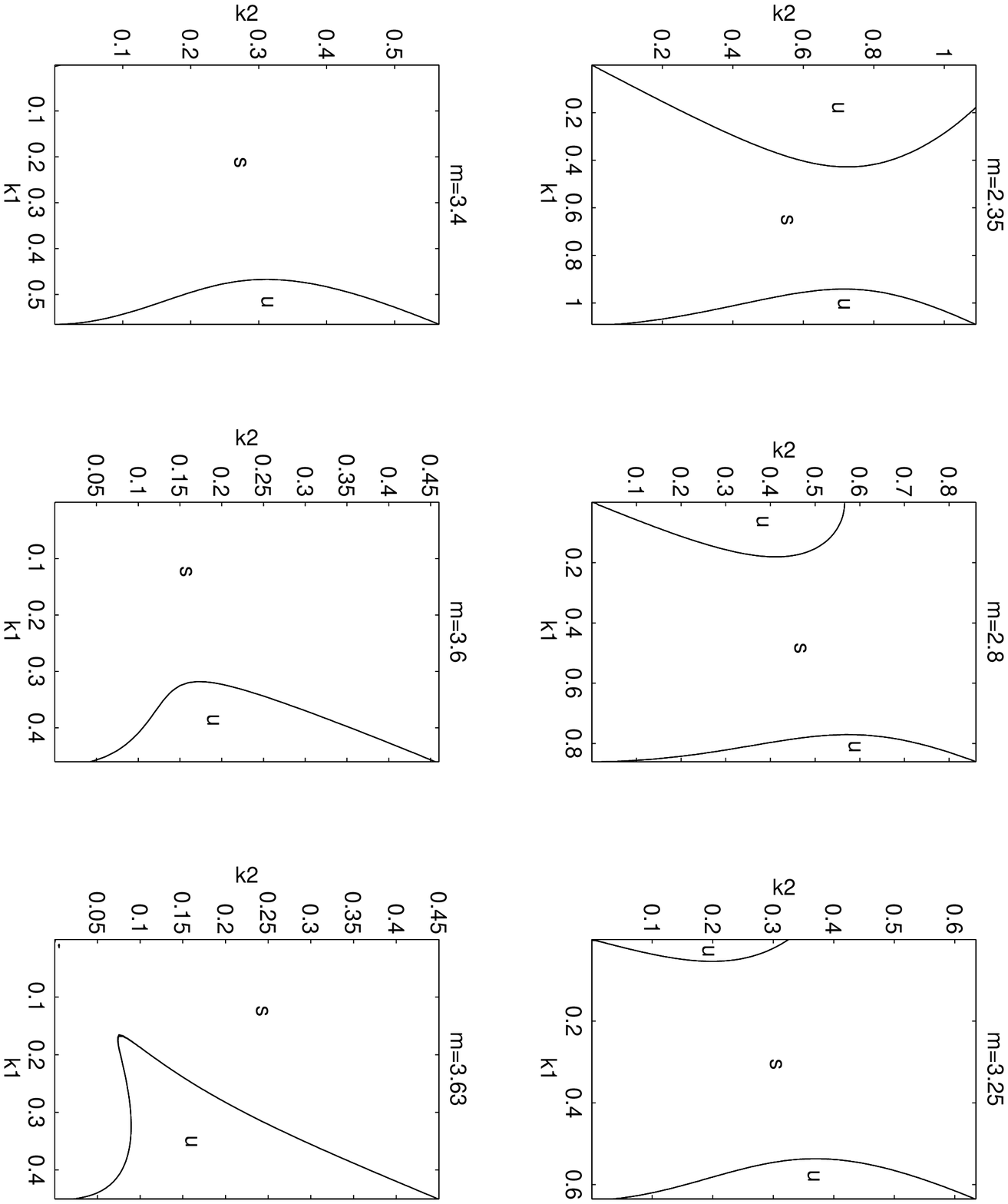}}
\end{picture}
\end{figure}

\newpage
\begin{figure}[htb]
\begin{picture}(420,240)(0,0)
\put(-100,-500) {\includegraphics{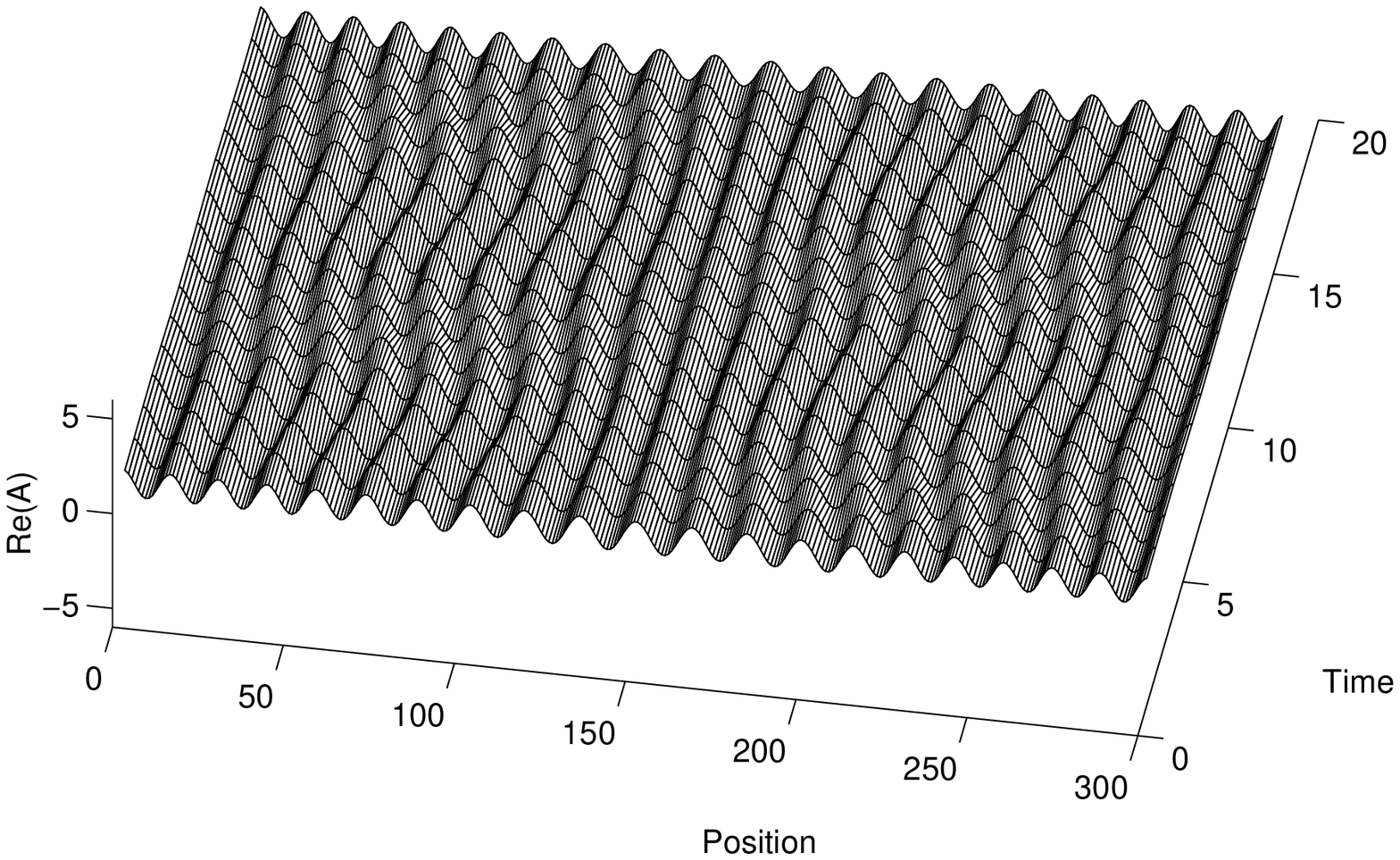}}
\end{picture}
\end{figure}

\newpage
\begin{figure}[htb]
\begin{picture}(420,240)(0,0)
\put(-100,-500) {\includegraphics{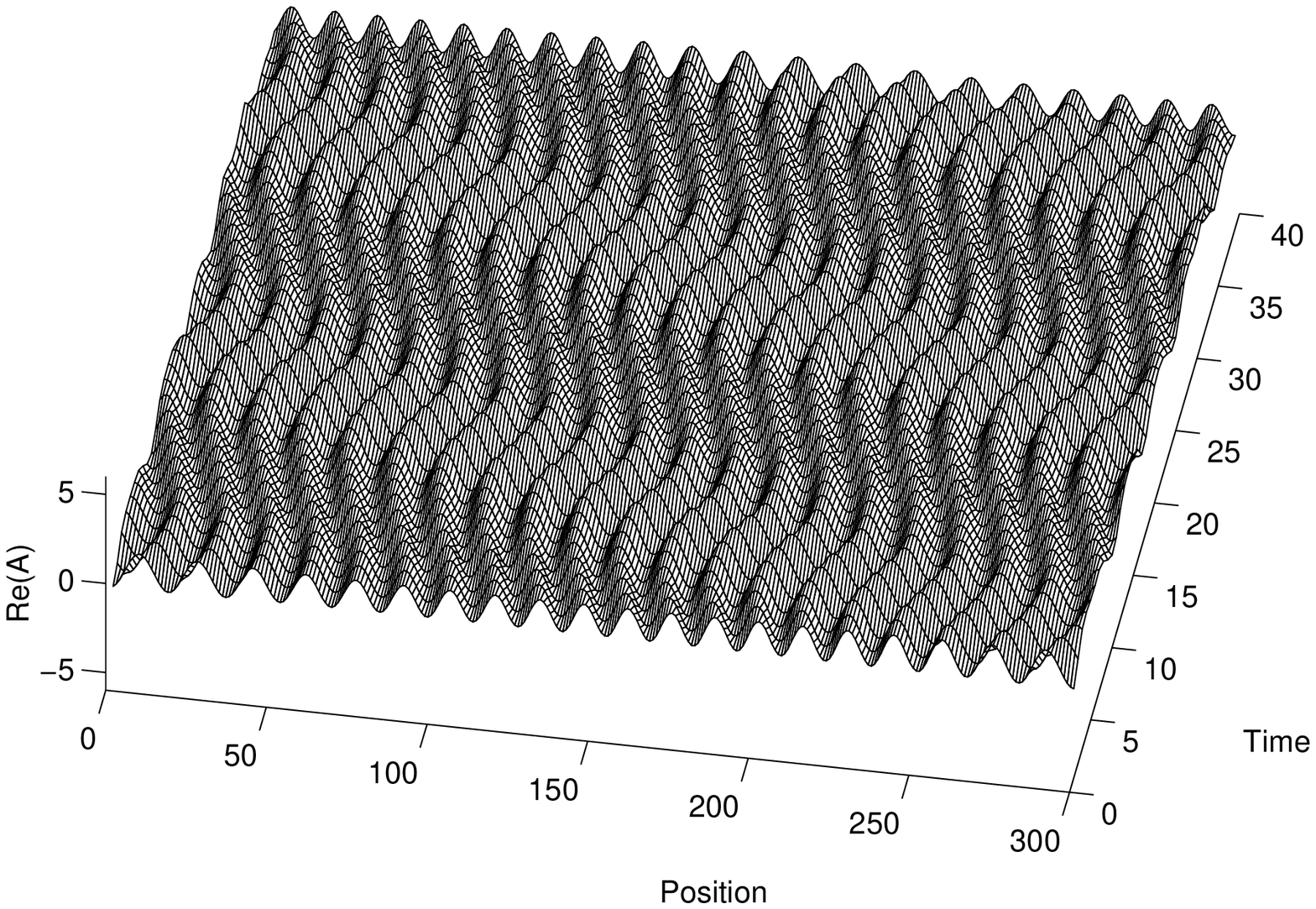}}
\end{picture}
\end{figure}

\end{document}